# TOAST Framework: A Multidimensional Approach to Ethical and Sustainable AI Integration in Organizations

Dian Tjondronegoro
Griffith University, Brisbane, Australia
d.tjondronegoro@griffith.edu.au

**Abstract**
Artificial Intelligence (AI) has emerged as a transformative technology with the potential to revolutionize various sectors, from healthcare to finance, education, and beyond. However, successfully implementing AI systems remains a complex challenge, requiring a comprehensive and methodologically sound framework. This paper contributes to this challenge by introducing the Trustworthy, Optimized, Adaptable, and Socio-Technologically harmonious (TOAST) framework. It draws on insights from various disciplines to align technical strategy with ethical values, societal responsibilities, and innovation aspirations. The TOAST framework is a novel approach designed to guide the implementation of AI systems, focusing on reliability, accountability, technical advancement, adaptability, and socio-technical harmony. By grounding the TOAST framework in healthcare case studies, this paper provides a robust evaluation of its practicality and theoretical soundness in addressing operational, ethical, and regulatory challenges in high-stakes environments, demonstrating how adaptable AI systems can enhance institutional efficiency, mitigate risks like bias and data privacy, and offer a replicable model for other sectors requiring ethically aligned and efficient AI integration.

## 1. Introduction

Artificial Intelligence (AI) significantly transforms industries and drives innovation in the digital age. Technological advancements and increased global investments have accelerated the adoption of AI in business operations. The adoption rates of AI range between 50-60% (Maslej et al., 2023), a significant increase compared to 2017, when only 20% of enterprises had incorporated AI into their workflows (Ransbotham, 2017). Organizations leverage AI to gain competitive advantages, create new market opportunities, and improve product quality and processes. Despite the increasing use of AI in various industries, many organizations face challenges when effectively integrating AI systems. Implementing AI without thorough planning and stakeholder alignment can lead to rushed decisions, wasted resources, and subpar outcomes (Li et al., 2023).

The rapid growth of AI automation brings forth issues such as job displacement, the need for reskilling, and increased responsibility for data privacy and security (Borges et al., 2021; Enholm et al., 2022). Additionally, concerns regarding regulatory compliance, ethical considerations, data quality, organizational readiness, and insufficient training and support can hinder the effective implementation of AI (Ganapathi & Duggal, 2023). Therefore, taking a strategic and responsible approach to AI implementation is crucial to ensure it aligns with organizational objectives and ethical standards (Sartori & Theodorou, 2022). Recent studies highlight the crucial role of senior leaders in ensuring that AI implementation aligns with the organization's overall objectives. Strategic alignment and leadership are emphasized in this process (Abonamah & Abdelhamid, 2024), and a new diagnostic framework has been proposed to tackle the challenges of integrating AI. It emphasizes the significance of building a network of partners and establishing an ecosystem for creating value (Ångström et al., 2023). Trust is the essential foundation to enable the value creation and foster innovation, creating a positive cycle where trust leads to innovation, and vice versa, through responsible AI practices.

The guiding question for our research is: "How can the development of adaptive AI systems incrementally build trust and integrate effectively into organizational structures while ensuring adherence to ethical standards, promoting a culture of innovation, and achieving socio-technical integration?". This paper contributes to the existing knowledge by proposing the TOAST framework, designed to guide organizations through the complexities of AI integration, ensuring that AI initiatives are ethically aligned, socially responsible, and conducive to continuous innovation. The TOAST framework emphasizes that successful AI implementation requires a deep understanding of AI technologies, their strategic roles, and their impact on organizational performance and societal well-being. It addresses challenges faced by internal and external stakeholders, including workforce implications, decision support, ethical considerations, and broader ethical, legal, and societal impacts.

Focusing on healthcare case studies provides a compelling context for examining the TOAST framework's relevance and utility. Healthcare settings represent some of the most complex environments for AI adoption due to high-stakes decision-making, rigorous regulatory standards, and deeply rooted organizational cultures. AI in healthcare demands precise integration to enhance clinical decision-making, optimize operations, and manage sensitive patient data responsibly. Case studies offer valuable insights into these challenges, highlighting the unique barriers and opportunities for implementing AI systems that are both technically robust and ethically sound. Moreover, the healthcare sector's rapid digital transformation, accelerated by the COVID-19 pandemic, underscores the need for adaptive, transparent, and trustworthy AI systems that align with organizational goals while upholding public trust. By grounding the research in healthcare, this paper draws from real-world complexities to explore how the TOAST framework can enable AI solutions that are scalable, socially responsible, and aligned with the high standards required in critical care settings.

The novelty of this paper lies in its interdisciplinary approach to AI integration, drawing from management, computer science, ethics, and social sciences to create a robust, adaptable framework. The TOAST framework's emphasis on trust-building and adaptability within the socio-technical environment positions it as a versatile model that can transcend healthcare to support AI integration across diverse industries. By prioritizing ethical governance, system optimization, and alignment with organizational culture, the framework establishes a standard for sustainable AI implementation that addresses both technical and human-centric requirements. This adaptability allows the TOAST framework to meet the distinct needs of sectors like finance, manufacturing, and education, where AI's integration similarly demands transparency, ethical compliance, and alignment with specific organizational structures. As such, TOAST not only sets a new benchmark for trust and adaptability in healthcare but offers a replicable model for other industries, enabling a consistent approach to AI deployment that balances innovation with responsible practices.

This paper is structured as follows. Firstly, we comprehensively analyze the socio-technical challenges associated with AI deployment. Secondly, we propose the TOAST Framework as a multidisciplinary guide that encompasses ethical, organizational, operational, and reputational dimensions of AI implementation, ensuring AI initiatives are ethically aligned, socially responsible, and innovative. Thirdly, we demonstrate the TOAST framework's practical and theoretical soundness through healthcare case studies, showing its applicability in real-world scenarios.

## 2. Literature Review

Integrating Artificial Intelligence (AI) within organizational frameworks presents a multifaceted challenge that necessitates a deep dive into social and technical dimensions. This section will begin by discussing the challenges of AI implementation as identified in the literature. It will then review the existing frameworks for AI implementation and identify the gaps the TOAST framework aims to fill. Finally, it will review the case studies selected from the literature to illustrate the challenges of AI implementation and justify the need for a comprehensive framework like TOAST.

### 2.1. AI Implementation Challenges

The increasing dependence on AI systems presents significant challenges that need thorough consideration, especially in decision-making roles (Loureiro et al., 2021). The 'black box' nature of AI systems poses a challenge to trust development, leading to skepticism and reluctance, particularly when the decisions have substantial impacts. Therefore, there is a growing call for explainable AI to communicate its reasoning to users (Adadi & Berrada, 2018). Enhancing transparency and explainability is fundamental for fostering responsible AI usage (Duan et al., 2019; Loureiro et al., 2021). Control over AI systems is closely related to trust, as it involves the ability to comprehend, predict, and influence AI behavior (Hevner & Storey, 2023). The challenge is to create systems that balance autonomy with control, ensuring AI operates under human supervision and aligns with organizational and ethical standards (Sartori & Theodorou, 2022).

Integrating AI in organizations demands a holistic approach considering the socio-technical integration between AI technology, organizational culture, workforce dynamics, and strategic alignment (Dwivedi et al., 2021; Loureiro et al., 2021). Visionary leaders are essential for integrating AI into organizational culture and processes, leveraging its strengths, and minimizing risks. (Cao, 2023; Chowdhury et al., 2022; Makarius et al., 2020). The transition to AI-enhanced workplaces must be managed responsibly, positioning AI as a collaborator that enhances human potential and addresses the need for reskilling and cultural readiness (Chowdhury et al., 2022). The success of AI adoption depends on aligning AI initiatives with strategic goals, the organization's cultural preparedness, and integrating AI into existing workflows (Uren & Edwards, 2023). It is essential to evaluate AI's impact on economic productivity and the trade-offs of different policy approaches for its adoption to validate the productivity advantages where AI supports the human workforce (Furman & Seamans, 2019).

AI offers significant opportunities for innovation but also requires substantial investments in change management and strategic alignment (Burström et al., 2021). Innovative business models incorporating AI must adhere to ethical standards and sustainable competitive practices (Attard-Frost et al., 2023). A responsible approach to innovation, informed by an understanding of AI deployment's ethical and societal implications, is imperative for sustaining public trust and the acceptance of AI technologies (Alhashmi et al., 2020; Arias-Pérez & Vélez-Jaramillo, 2022).

## 2.2. Existing AI Implementation Frameworks

A key principle that underpins implementation frameworks is *Responsible AI*, "the practice of developing, using, and governing AI in a human-centered way to ensure that AI is worthy of being trusted and adheres to fundamental human values" (Vassilakopoulou, 2020). It signifies the development of intelligent systems that maintain fundamental human values to ensure human flourishing and well-being in a sustainable world (Dignum, 2019). AI systems should also contribute toward global sustainability challenges by ensuring effective computational models through energy-aware solutions and greener data centers and promoting AI use to help achieve sustainability goals (Chatterjee & Rao, 2020). Business models for responsible AI can be developed by creating new or specific business models, using ethical principles for navigating the potential conflicts between commercial and societal interests, and emphasizing the criticality of social responsibility (Zimmer et al., 2022). The four pillars of Responsible AI are organizational, operational, technical, and reputational (Eitel-Porter et al., 2021). These pillars schematize the distinct yet interconnected guiding principles that guide organizations through the various stages of AI maturity, ensuring a balanced progression that aligns with ethical standards, operational integrity, and stakeholder trust.

*Dynamic Trust Framework* emphasizes that trust in AI is not static; it evolves with technology, user experiences, and societal perceptions (Adewuyi et al., 2019). The framework includes mechanisms for the dynamic and incremental assessment of trust in AI systems (Cabiddu et al., 2022). Dynamic trust involves regular evaluation of user experiences, transparency in AI operations, and responsiveness to stakeholder concerns (Robinson, 2020). *AI Accountability Framework* is pivotal in the governance of AI systems, supported through key performance indicators and internal auditing that measure both ethical and operational aspects of AI (Tóth et al., 2022). These metrics provide a quantifiable means to assess compliance with ethical standards, effectiveness, efficiency, and overall (end-to-end) impact of AI systems (Raji et al., 2020). The *TOE Framework* assesses technological, organizational, and environmental factors in AI adoption at different stages. It focuses on integrating advanced and ethical AI technologies, establishing clear roles and governance structures, maintaining regulatory compliance, leveraging external resources effectively, and upholding the organization's reputation and trust in its AI technologies (Neumann et al., 2022).

*The Ethical AI Framework* prioritizes continuous ethical investigations to assess moral implications across all aspects of Responsible AI, including AI decision-making, algorithm fairness, and user privacy and rights. (Prem, 2023). It ensures that AI systems align with societal values and ethical norms by embedding them within the organizational culture, operational processes, technical development, and reputation management (Floridi et al., 2021). *AI Risk Framework* focuses on identifying, analyzing, and mitigating potential risks associated with AI systems. It involves a proactive approach toward recognizing potential threats, from data privacy breaches to biased decision-making, and establishing protocols to prevent or minimize their impacts (Wirtz et al., 2022). It emphasizes risk mitigation and the maximization of positive impacts, ensuring that AI systems contribute constructively to organizational goals and societal welfare (Tabassi, 2023).

## 2.3. Case Studies in AI Integration into Organizations

Focusing on healthcare case studies for AI integration offers a critical contribution to the current literature due to the distinct challenges and transformative potential AI holds within this sector. Healthcare systems are complex, involving high-stakes decision-making, regulatory constraints, and deeply embedded workflows that necessitate specialized approaches to technology adoption. Case studies provide nuanced insights into how AI can address these complexities—enhancing clinical decision-making, streamlining operations, and refining diagnostics—all while navigating ethical, data privacy, and interoperability concerns. By examining real-world applications, healthcare case studies contribute a grounded perspective on both the operational and ethical frameworks needed to implement AI responsibly, helping to bridge gaps between theoretical AI models and practical, scalable solutions that enhance patient care and institutional efficiency.

Seven healthcare-related case studies from the literature have been selected for our analysis. Each study reveals specific challenges, diverse perspectives, and findings, offering a multidimensional view of AI implementation across this sector.

(Ganapathi & Duggal, 2023): This study delves into the challenges and opportunities associated with AI implementation in the UK's National Health Services (NHS) from practitioners' perspective. Through thematic analysis of eleven semi-structured interviews with doctors, it was found that the main hurdles include the need for structured pathways into the AI field, a steep learning curve, and greater comfort with uncertainty. The study emphasizes the importance of involving doctors in developing AI tools to leverage their clinical knowledge for better oversight and effective AI integration.

(Morrison, 2021): Morrison's research applies innovation theory to explore solutions to AI adoption barriers within the NHS, focusing on radiology, pathology, and general practice. Twelve interviews with key informants, including NHS doctors, managers, and regulatory personnel, revealed that IT infrastructure quality and financial pressures are significant barriers. The study highlights the need for clear language around AI and establishing a gold standard for AI deployment.

(Petersson et al., 2022): This study investigates healthcare leaders' perspectives on AI implementation in Swedish healthcare. Through qualitative content analysis of 26 interviews, it identifies the transformative impact of AI on healthcare professions. Challenges beyond the healthcare system's direct control, such as strategic change management, are significant. The study emphasizes the need for systematic approaches and shared plans level to address these challenges, as well as support from top leadership for successful AI adoption.

(Li et al., 2023): This study explores how AI transforms healthcare's Human Resource Management (HRM). Their interviews with HRM staff and analysis of archival data highlight that AI facilitates co-creation processes and personalized care pathways and improves performance by cutting costs and extending service offerings. The study underscores AI's potential to revolutionize HRM through targeted, individualized approaches to patient care.

(Dumbach et al., 2021): This cross-national comparison examines AI adoption in healthcare SMEs in China and Germany, highlighting cultural and regulatory differences and their impact on AI development. Fourteen semi-structured interviews with managers from both countries reveal that AI is increasingly used to enhance innovation and product development, including medical image diagnosis and autonomous robotic surgery. The study suggests that SMEs adopt AI to co-create processes and build personalized care pathways, although regulatory and privacy issues need addressing.

(Henry et al., 2022): This study explores clinicians' experiences with a deployed machine learning (ML) system in a clinical setting, focusing on trust and human-machine teaming. Through interviews, observations, and experimental studies with 20 clinicians using a "targeted real-time warning" system, the study finds that clinicians see ML systems as supportive tools in diagnosis and treatment. However, concerns about potential over-reliance on automated systems could degrade clinical skills over time.

(Sun, 2021): Sun's research investigates the interplay of technology, power, and organizational behavior in AI adoption in Chinese hospitals. The multi-case study approach and 29 interviews examine how stakeholders use their social power to influence AI adoption. Challenges include insufficient data quality for training AI systems and the necessity for developed AI systems to be introduced in hospitals.

The inclusion criteria for these healthcare case studies focused on capturing a comprehensive view of AI implementation challenges, aligned with the TOAST framework's core components: trust, optimization, adaptability, and socio-technical integration. Each study, selected for its relevance to different healthcare applications and geographic contexts, addresses critical challenges such as regulatory barriers, IT infrastructure, data privacy, and impacts on professional roles. Ganapathi & Duggal (2023) and Morrison (2021) provide insights into structural and logistical challenges within an organization (the NHS), while Petersson et al. (2022) and Henry et al. (2022) explore the importance of clinician trust and acceptance in clinical AI deployment. Li et al. (2023) and Dumbach et al. (2021) expand the focus to non-clinical applications and SMEs, emphasizing adaptability and regulatory compliance across diverse settings. Sun (2021) highlights socio-political dynamics in Chinese hospitals, focusing on power relations and data quality issues. These selected case studies offer a multidimensional basis for evaluating the TOAST framework in real-world settings.

## 3. Methodology

Our research is based on a comprehensive literature review and case studies analysis on AI implementation's ethical, socially responsible, and innovation-oriented impacts. We integrated the existing multidisciplinary theories to develop the TOAST framework as a unifying view for responsible AI use and management. We then applied the framework on the selected case studies in the healthcare sector to evaluate the theoretical soundness and deepen our understanding of technology, ethics, and operational demands.

The **literature review methodology** focused on exploring and integrating multidisciplinary theories and frameworks across different AI maturity levels within organizations, acknowledging each level's unique challenges and needs. We searched academic journals from databases such as Google Scholar, IEEE Xplore, ACM, and AIS eLibrary. The search keywords include "organizational AI implementation," "multidisciplinary AI frameworks," and "responsible AI." We have reviewed literature from the last ten years to ensure relevance and recency in the context of rapid technological advancements. Non-English studies and those focused solely on the technical aspects of AI were excluded to maintain a clear organizational and implementation perspective.

Through a rigorous thematic analysis, we synthesized data from diverse studies, distilling essential concepts such as trust, control, socio-technical alignment, and innovation. This method enabled us to identify and integrate consistent patterns and principles across disciplines, clarifying the shared challenges that effective AI implementation frameworks must address. The TOAST framework is directly grounded in these findings, blending insights from management, ethics, and technology fields to define the critical components of a robust AI implementation strategy. By anchoring each of TOAST's core elements—trust and accountability, optimization and control, adaptability and innovation, and socio-technical harmony—in these cross-disciplinary insights, we crafted a framework that is both theoretically rigorous. These components collectively address the complex needs outlined in the literature, ensuring that TOAST offers a comprehensive, adaptable model for guiding AI integration across various organizational environments.

The **case study methodology** was designed to ground the TOAST framework in real-world and practical cases, ensuring its theoretical soundness and practical applicability. In the initial stage, we carefully selected diverse case studies from the literature, focusing on those that offered insights into real-world AI adoption challenges, methodologies, and varying levels of AI maturity. Each case was chosen for its empirical depth, typically gathered through interviews, providing practical perspectives on the complexities of integrating AI in healthcare. Additionally, each study included a robust literature review and theoretical analysis to anchor its findings in broader AI implementation theory, ensuring that our framework builds upon validated insights.

The second stage involved detailed data extraction from each case study, capturing a wide array of information, including specific AI implementation strategies, organizational challenges, and the solutions and outcomes observed. This data provided a rich foundation for understanding the situational factors influencing AI adoption, such as regulatory requirements, technical constraints, and workforce considerations. By examining these factors, we were able to assess how they impact each stage of AI maturity, offering a comprehensive view of the practical challenges and strategies involved in AI integration. In the third stage, we conducted a comparative analysis across the selected cases, identifying common themes, critical differences, and nuanced insights into AI adoption. By systematically contrasting each case, we discerned patterns in how organizations address trust, control, socio-technical integration, and innovation challenges. This comparative approach provided a cohesive understanding of how AI implementation strategies and outcomes vary across organizational contexts, contributing to a nuanced validation of the TOAST framework's principles and its adaptability across different stages of AI maturity in healthcare.

## 4. Framework Design

This section explores the theoretical foundations and hypotheses essential for AI implementation in organizational contexts. By examining various theoretical perspectives, we designed a robust framework to understand AI's multifaceted impact, emphasizing trust, accountability, optimization, adaptability, and socio-technical alignment. The framework bridges the gap between theory and practice, offering a structured approach for sustainable and effective AI integration in organizational settings.

## 4.1. Discussion of Key Theories

*Trust theory*, examined across disciplines like psychology, sociology, and computer science, underscores the importance of transparency and ethical considerations in building user confidence in AI systems. AI systems must be perceived as useful, reliable, and influenced by social factors to gain user acceptance (Cho et al., 2015; Dumbach et al., 2021; Gille et al., 2020; Omrani et al., 2022; Vereschak et al., 2021). Trust is dynamic and context-dependent, requiring AI systems to be adaptable to various domains and user needs, ensuring ethical, transparent operations aligned with human values (Araujo et al., 2020; Ferrario et al., 2020; Glikson & Woolley, 2020; Jacovi et al., 2021; Razmerita et al., 2022).

*Control theory*, rooted in engineering and mathematics, leverages algorithms and feedback mechanisms to manage system behavior, allowing for self-adjustment and optimization in response to new data (Filieri et al., 2015; Gill et al., 2022; Hou & Wang, 2013). This theory's application in AI advocates for a symbiotic relationship between human intuition and AI's computational autonomy, contributing to organizational learning and sustainability (Bankins et al., 2023; Caldas et al., 2020; Khargonekar & Dahleh, 2018; Seok et al., 2012).

*Socio-technical theory* advocates for culturally compatible AI ethically aligned with stakeholder values. This approach ensures that AI systems are accountable by considering privacy, transparency, and fairness as integral components (Baxter & Sommerville, 2011; Hoda, 2022; Holton & Boyd, 2021; Sartori & Theodorou, 2022; Toreini et al., 2022; Vassilakopoulou, 2020).

*Innovation theory* positions AI technologies as transformative agents that fundamentally shift innovation trajectories across various domains and industries. It includes the principles for grasping AI's role in instigating systemic changes in business models, organizational strategies, and market dynamics (Haefner et al., 2021; Mariani et al., 2023), underscoring organizations' need to evolve to stay relevant and competitive (Burström et al., 2021; Di Vaio et al., 2020). AI cultivates innovative products, services, processes, and business practices that often represent significant advancements, redefining industry benchmarks and consumer expectations (Alsheibani et al., 2018; Jöhnk et al., 2021; Verganti et al., 2020).

## 4.2. Synthesis of Theories Linked to Hypotheses

*Hypothesis 1 (H1): Development of Adaptive AI Systems Cultivates Incremental Trust*

Establishing trust in AI systems is a gradual process that strengthens over time. Trust develops as users engage with the system and observe its consistent reliability and effectiveness. The context in which AI software is deployed—including the user environment, application specificity, and the domain of use—significantly influences trust formation (Ferrario et al., 2020). AI software engineering must address socio-technical complexities to build trust by integrating AI into existing systems and aligning it with user expectations (Cho et al., 2015; Gille et al., 2020) while incorporating technical soundness, social implications, ethical guidelines, user experience, and societal impact (Baxter & Sommerville, 2011; Hoda, 2022).

AI systems must adapt dynamically by observing and adjusting their behavior to meet objectives. Control theory has been used in software engineering to create self-adjusting software systems. It is important to incorporate feedback control strategies and strong adaptation principles to improve AI systems' resilience in unpredictable conditions(Caldas et al., 2020; Filieri et al., 2015). Innovation theory is also relevant to AI development, from business model innovation to practical AI implementation. A comprehensive approach to AI adoption is required, balancing technological components with human factors in the organization (Burström et al., 2021; Mariani et al., 2023).

*Hypothesis 2 (H2): Transparent AI Algorithms and Models Enhance Human-centered Accountability and AI Readiness*

The cultivation of trust among users requires transparency in AI algorithms and models, along with ethical considerations to ensure impartiality and absence of bias (Liu, 2021). The type of tasks executed by AI algorithms profoundly influences trust levels. Different types of tasks influence trust levels, with technical tasks often garnering more trust than those involving social intelligence or ethical discernment (Glikson & Woolley, 2020). Control theory is crucial for designing and refining AI algorithms, enabling real-time performance enhancement, system resilience (Hou & Wang, 2013) and system performance optimization, especially in real-time settings (Mueller et al., 2019).

Socio-technical theory emphasizes the need for accountable and transparent AI systems that prioritize human values, clarity, and equity while addressing issues related to bias, privacy, and ethical application (Herrmann & Pfeiffer, 2023; Vassilakopoulou, 2020). Innovation theory highlights the importance of employees' awareness of AI and organizational readiness for AI integration (Arias-Pérez & Vélez-Jaramillo, 2022; Jöhnk et al., 2021). A holistic approach to AI adoption within organizations, including understanding employees' perspectives on AI and ensuring strategic preparedness, is essential for successful AI integration (Brock & Von Wangenheim, 2019).

*Hypothesis 3 (H3): Responsible AI Innovation Sustains Societal Trust in AI*

Establishing trust in artificial intelligence (AI) requires a responsible innovation strategy to ensure equitable benefits. This involves accountability, privacy, fairness, explainability, transparency, reproducibility, reliability, user interaction, data safeguarding, and security threat mitigation (Chatila et al., 2021; Cho et al., 2015). Adapting to AI technology challenges involves balancing data sharing for training while protecting privacy and addressing safety, privacy, and bias concerns. AI systems must also consider social, economic, and environmental sustainability, as well as information security control and ethical considerations in data sharing (Anderson et al., 2017; Seok et al., 2012).

The societal discourse and perceptions of AI are crucial in shaping its development and public image. Engaging the public is important for establishing a positive relationship with AI (Sartori & Bocca, 2023; Sartori & Theodorou, 2022). Using non-explainable AI models requires balancing performance advantages and associated risks (Asatiani et al., 2021). Incorporating AI algorithms into business operations can enhance efficiency and lead to innovations in products and services, improving decision-making capabilities and business models(Burström et al., 2021; Haefner et al., 2021). An organization's readiness for AI innovation depends on aligning AI technologies with strategic objectives, resource availability, and fostering an innovation-friendly culture (Alsheibani et al., 2018; Arias-Pérez & Vélez-Jaramillo, 2022; Jöhnk et al., 2021).

*Hypothesis 4 (H4): Human-AI Symbiotic and Synergistic Collaborations Enable Transformational Impact*

Trust in AI is crucial for effective collaboration and decision-making in autonomous systems, especially when AI's decisions have significant consequences for users (Jacovi et al., 2021). Human-AI interactions are pivotal in cultivating trust within organizational contexts and are influenced by individual user characteristics, such as AI literacy, privacy concerns, and cultural diversity (Araujo et al., 2020). Trust is contingent upon the degree of autonomy granted to AI and its effects on users (Liu, 2021). Human-AI symbiosis suggests that AI systems should enhance human cognitive capabilities, and control theory emphasizes the need for psychologically and technically proficient systems (Clemmensen, 2021). Human and AI interactions within organizations reveal differing perceptions of competencies, employee attitudes toward AI, and the broader implications of AI on the labor market (Bankins et al., 2023; Seok et al., 2012). AI's capability to process complex datasets fosters a symbiotic relationship that enhances organizational decision-making (Jarrahi, 2018; Soma et al., 2022). Proposals for advancing human-centered AI include a two-dimensional framework for high levels of human control and automation, redefining AI systems as potent tools under human command, and a tripartite governance structure to aid software engineering teams in developing reliable, safe, and trustworthy systems (Shneiderman, 2020).

The integration of social and technical systems is crucial for the trustworthy implementation of AI technologies. It emphasizes transparency, human oversight, and equitable deployment, shaping employment trends, job dynamics, and organizational roles (Chowdhury et al., 2022; Sartori & Theodorou, 2022; Yu et al., 2022), Integrating AI requires knowledge dissemination, workplace AI socialization, management of evolving AI systems, and adjustments to decision-making frameworks (Herrmann & Pfeiffer, 2023; Sony & Naik, 2020). AI innovation has substantial transformative potential, enhancing service provision, decision-making, and operational efficiency. For example, AI has improved patient outcomes and reduced healthcare costs (Alhashmi et al., 2020). It also influences job satisfaction, psychological well-being, and organizational productivity (Yu et al., 2022). However, the integration of AI requires careful consideration of data privacy, ethical imperatives, and human-centric services. The growing prevalence of autonomous systems demands a nuanced comprehension of the dynamics between humans and AI (Verganti et al., 2020).

## 4.3. The TOAST Framework as a Unifying View

By melding multidisciplinary theories, the proposed framework provides an enriched, flexible, and adaptable blueprint for AI's responsible use and management as augmented intelligence. It aims to guide organizations through the complexities of AI implementation, ensuring ethical responsibility, social harmony, and continuous innovation. Figure 1 summarizes the proposed framework, highlighting the key components and how they interrelate and synergize.

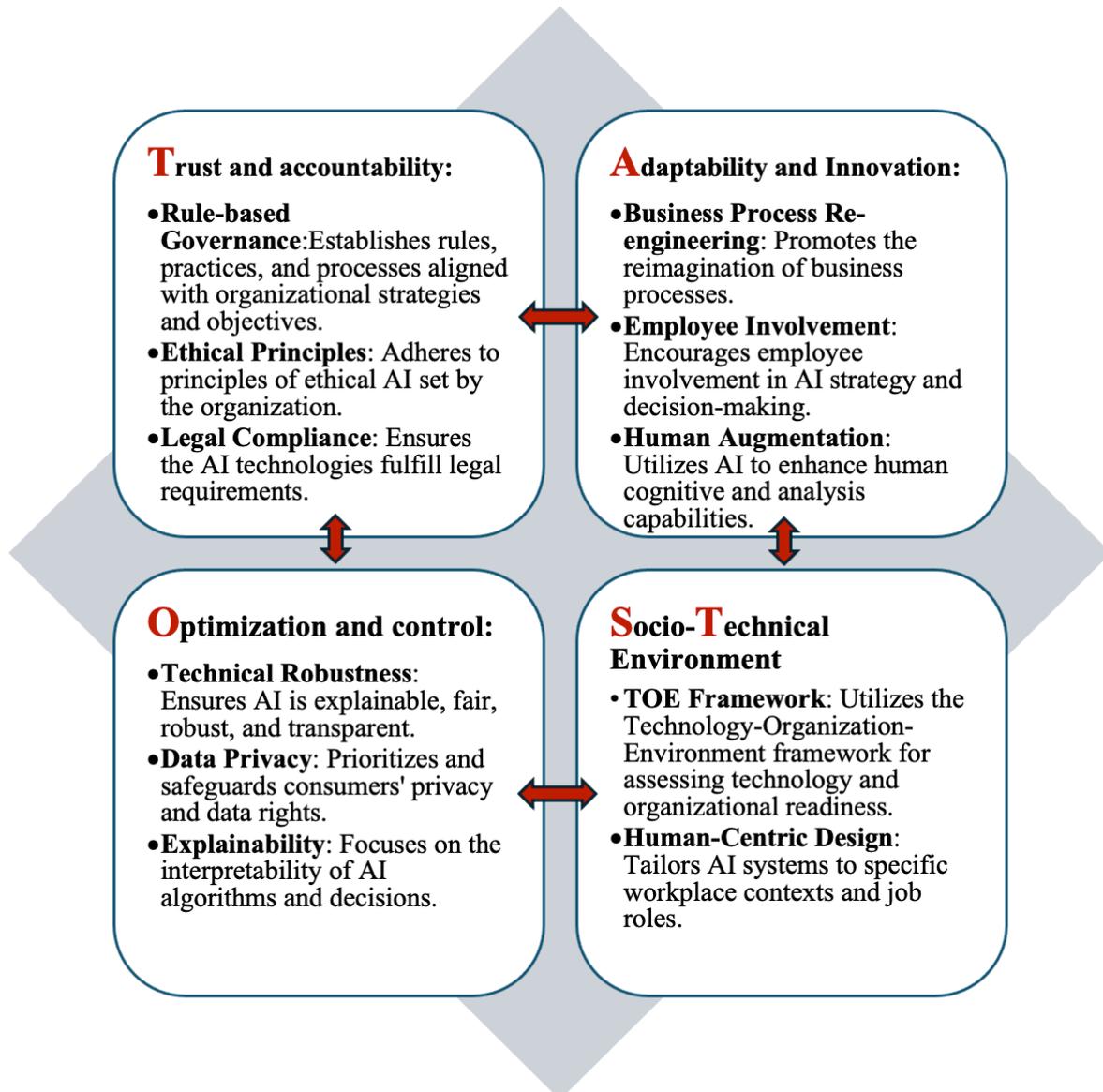

Figure 1. Proposed TOAST Framework

*T - Trust and accountability (Governance and Ethics)*: This component is the ethical compass of the framework, anchored in trust theory. It prescribes comprehensive guidelines for instilling and sustaining trust in AI systems through clear governance, ethical practice, and legal compliance. We advocate layered trust mechanisms for governing AI systems. At the first (internal) layer, AI system should be transparent and explainable so that people can understand how they make decisions and ensure that AI's use in business operations is aligned with organizational values and strategies. The next (external) layer is a strict adherence to legal standards and societal values, applying governance principles to ensure that AI technologies are ethical and legally compliant. Adaptive governance combines global and context-specific ethical principles, reflecting the organization's commitment to uphold ethical AI practices.

*O - Optimization and control (Technical Requirements):* This component is the technical backbone of our framework, drawing inspiration from control theory. To achieve a user-centered system, it aims to optimize the performance of AI systems within organizational workflows through meticulous calibration, responsive feedback mechanisms, and adaptive tuning to meet evolving operational needs. By prioritizing technical robustness principles, the component ensures that AI systems are efficient, transparent, and fair. It also emphasizes the importance of data privacy, consumer data rights protection, and the explainability and interpretability of AI algorithms and decisions.

*A - Adaptability and innovation (Human-AI Collaboration):* This component fosters continuous innovation and harnesses AI's transformative potential by promoting adaptable, user-centered systems that evolve with organizational needs. Aligned with innovation theory, it emphasizes collaboration between human expertise and AI, ensuring solutions are technically advanced yet adaptable to new challenges and opportunities. By encouraging re-engineering of workflows and a fresh perspective on business models, this approach positions AI as a tool for enhancing decision-making and expanding cognitive capacities. Active employee involvement is crucial, creating an environment where staff contribute to AI strategy and decision-making, fully embracing human augmentation to enhance both cognitive and operational capabilities.

*ST - Socio-Technical Harmony (Organizational Readiness):* Informed by socio-technical theory, this component recognizes that AI is deeply interconnected with organizational structures, culture, and human factors. Successful AI implementation requires aligning technology with users' skills, workflows, and values, creating systems that enhance rather than replace human capabilities. By embedding AI within the broader socio-technical landscape, this approach promotes resilient, adaptable, and seamlessly integrated solutions that fit organizational needs and elevate user engagement and skill. Frameworks like TOE assess both technological and organizational readiness, ensuring AI systems are customized, user-centered, and well-integrated into the workplace.

## 5. Case Study Analysis

This section will apply the TOAST framework to the selected case studies, providing a robust evaluation of its practicality and theoretical soundness in diverse healthcare settings. By systematically analyzing each case, we will assess how well the framework addresses real-world challenges such as trust-building, technical optimization, adaptability, and socio-technical integration. This application will highlight the framework's ability to guide AI implementation across varying stages of maturity and organizational contexts, demonstrating its adaptability and relevance beyond theoretical constructs.

*Trust and Accountability (T):* Accountable AI implementation in healthcare hinges on transparent, ethically governed systems that foster trust across contexts. Ganapathi and Duggal (2023) and Henry et al. (2022) emphasize transparency from distinct angles: Ganapathi and Duggal focus on governance structures to sustain NHS doctors' trust, while Henry et al. add that clinician interaction with AI development teams enhances experiential trust, aligning with TOAST's emphasis on layered trust mechanisms. Petersson et al. (2022) and Li et al. (2023) highlight accountability from a policy perspective; Petersson advocates for robust legal frameworks, whereas Li et al. pushes for policies that both regulate and actively promote digital healthcare, reflecting TOAST's adaptive governance goals. Dumbach et al. (2021) and Sun (2021) extend the discussion to technical and socio-political trust barriers, such as data privacy and power dynamics, arguing that trust-building in cross-national SME contexts requires not only ethical governance but also rigorous protections and adaptive structures. Collectively, these perspectives underscore TOAST's commitment to ethically aligned, contextually responsive AI governance.

*Optimization and Control (O):* Effective AI implementation in healthcare requires robust technical infrastructure, stringent regulatory adherence, and precise optimization strategies to ensure seamless integration and control within complex clinical environments. Ganapathi and Duggal (2023) and Sun (2021) emphasize optimized calibration and tuning, aligning with TOAST's focus on technical robustness. In contrast, Morrison (2021) and Petersson et al. (2022) prioritize data governance and regulatory adherence, underscoring that robust infrastructure must also meet legal standards—an approach TOAST fully integrates. Li et al. (2023) and Henry et al. (2022) highlight AI's role in boosting efficiency and supporting clinicians, aligning with TOAST's commitment to reliable, user-centered systems. Extending this to SMEs, Dumbach et al. (2021) argue that quality and compliance are essential for sustainable AI innovation. Thus, TOAST's framework addresses these varied needs by ensuring that AI systems are not only technically and legally sound but also adaptable, efficient, and reliable across diverse healthcare settings.

*Adaptability and Innovation (A):* Adaptability and innovation are central to fostering human-AI collaboration, aligning with TOAST's focus on adaptive, transformative AI solutions that integrate seamlessly across organizational units. Ganapathi & Duggal (2023) and Henry et al. (2022) emphasize AI's need to adapt within clinical workflows and existing NHS structures, reflecting TOAST's emphasis on scalable, user-centered systems. Morrison (2021) and Petersson et al. (2022) highlight economic and strategic planning requirements for adaptability, suggesting that sustainable AI integration depends as much on financial viability as on organizational readiness. In the SME contexts, Dumbach et al. (2021) and Li et al. (2023) underscore AI's role in re-engineering processes and enabling co-creation, revealing the importance of cultural and process-level flexibility. Sun (2021) extends this view by arguing that AI must be adaptable to different learning capabilities within healthcare applications, reinforcing TOAST's advocacy for human-centered, context-sensitive solutions. Combining these perspectives support TOAST's goal of creating an innovative, human-augmented environment that actively involves employees in AI-driven decision-making.

*Socio-Technical (ST):* The TOAST framework's focus on socio-technical harmony underscores the interdependence between AI systems and organizational environments, aligning technology with culture, workflows, and job roles. Ganapathi and Duggal (2023) and Morrison (2021) highlight structural barriers, such as regulatory standards and data fragmentation, suggesting that AI must be adaptable to specific organizational and data requirements. Petersson et al. (2022) and Henry et al. (2022) emphasize that AI integration should support professional roles and clinical workflows, minimizing disruption and aligning with TOAST's commitment to user-centered design. Addressing cultural and regulatory differences, Dumbach et al. (2021) and Sun (2021) underscore the need for AI systems that consider both external compliance and internal power dynamics, ensuring seamless integration and reducing resistance. TOAST thus promotes AI solutions that are technically robust yet sensitive to socio-organizational contexts, enhancing organizational readiness and fostering user trust.

*T-O Synergy:* The relationship between optimized AI systems and trust reflects a fundamental synergy: technical robustness must be paired with ethical transparency to gain healthcare professionals' confidence. Henry et al. (2022) highlight that clinicians' trust is rooted in a clear understanding of AI mechanics, suggesting that optimization alone is insufficient; clinicians also need transparency to fully trust AI. This contrasts subtly with Morrison (2021), who stresses that the overall quality of technology infrastructure is what underpins trust, implying that reliable, well-maintained systems are foundational even before transparency is factored in. Ganapathi and Duggal (2023) bridge these perspectives by arguing that transparent AI is necessary to earn NHS doctors' trust, effectively linking ethical governance with technical optimization. Thus, TOAST's approach reinforces that trust in AI systems is not solely a byproduct of technical accuracy or infrastructure quality but arises from a balanced combination of ethical transparency, technical reliability, and user-centered design, all of which contribute to a trustworthy AI architecture in healthcare settings.

*A-ST Synergy:* The adaptability and innovation of AI in healthcare hinge on a well-aligned socio-technical environment, where technology and organizational culture co-evolve. This synergy allows AI systems to meet immediate needs while remaining flexible enough to address future challenges, supporting long-term relevance and resilience. Petersson et al. (2022) underscore that systematic change management and strategic planning are essential to effectively integrate AI, suggesting that organizational readiness must be as adaptable as the technology itself. In contrast, Sun (2021) contends that social power dynamics are central to AI adoption, proposing that adaptability depends not only on planning but also on navigating internal hierarchies and influence networks. This perspective emphasizes that AI implementation requires an environment receptive to organizational and technological shifts, especially when workforce skill levels and cultural dynamics vary. Integrating these views reinforce the TOAST framework's emphasis on creating a socio-technical harmony where innovation is sustainably supported by both adaptable technology and a conducive organizational setting that can evolve alongside AI advancements.

*O-ST Synergy:* The optimization of AI in healthcare is inherently tied to the socio-technical environment, where alignment with organizational workflows, staff skills, and cultural dynamics is essential for effectiveness. Morrison (2021) argues that successful AI deployment must integrate seamlessly with existing practices, particularly in healthcare settings where diagnostic and treatment processes hinge on the nuanced expertise of medical staff. This suggests that AI optimization cannot be a one-size-fits-all approach; it must be customized to align with specific professional competencies and patient care demands. Contrastingly, Dumbach et al. (2021) focus on broader regulatory and data privacy barriers, especially within SMEs, implying that even highly optimized AI can struggle to deliver value if external socio-technical constraints, such as data governance, are not met. Hence, socio-technical alignment extends beyond organizational culture to regulatory compliance and data security, emphasizing that AI systems must operate within the broader socio-political landscape. Ganapathi and Duggal (2023) bridge these views by stressing that technical challenges like system interoperability and data access are critical barriers within the NHS. Their perspective suggests that optimization is not solely an internal process but one that also requires robust external connectivity and information accessibility. Combining these insights reinforce the TOAST framework's stance that AI optimization must address both internal alignment with workflows and broader socio-technical constraints, creating a cohesive environment where AI can operate effectively and securely.

*T-A Synergy:* Trust serves as a catalyst for both organizational adaptability and AI-driven innovation, creating a reinforcing cycle where trusted systems encourage exploration and adaptation, thus spurring further innovation. Ganapathi and Duggal (2023) highlight that trust, especially when built through clinician involvement, is foundational to fostering adaptability in healthcare AI, as it ensures the technology meets practical clinical needs. This perspective underscores that trust hinges not only on system reliability but also on the perceived relevance of AI solutions to user roles. Henry et al. (2022) extend this idea, showing that clinicians' hands-on experience with AI and peer endorsements are powerful drivers of trust, suggesting that direct engagement with AI systems is as critical as technological design for successful integration. This contrasts slightly with Li et al. (2023), who, in the HRM context, emphasize the importance of trust among administrative and policy stakeholders to drive digital transformation. Here, trust must be built at a strategic level, highlighting that in non-clinical contexts, organizational commitment to AI's long-term value plays a central role in establishing trust. Integrating these perspectives reinforce TOAST's emphasis that trust is not merely a byproduct of system accuracy but is actively constructed through stakeholder involvement, role relevance, and strategic alignment, This approach underscores the importance of a multifaceted approach to trust-building in enabling adaptability and sustained AI innovation.

*T-O-A-ST synergy:* The integrated analysis from these case studies demonstrates that responsible AI practices must balance technical optimization with socio-technical adaptability, fostering an environment where AI can thrive as a trusted, innovative tool in healthcare. By focusing on these interconnected components, healthcare organizations can effectively integrate AI to enhance patient care and operational efficiency, ensuring that AI systems are technologically advanced, socially responsible, and ethically grounded. The proposed framework emphasizes responsible AI practices that are socially aware and designed to build trust through transparency and accountability. AI systems must be embedded within an organization's fabric, balancing technological advancement with human-centric design, ethical governance, and agile adaptability. The vision for AI in healthcare, is to realize AI as a tool for augmenting human capabilities, guided by a responsible governance model that adapts to technological and societal shifts.

*Alignment of TOAST Strategies and the Proposed Hypotheses*

The TOAST framework's hypotheses and proposed strategies align well with the case studies. This alignment demonstrates that the framework is theoretically sound and provides a robust foundation for AI implementation in healthcare.

Adaptive AI Systems Development (Hypothesis 1):
- Prediction: AI systems must be adaptable to integrate seamlessly into existing technical infrastructures and workflows.
- Outcome: Building incremental trust in the system through adaptability.
- Case Study Alignment: Ganapathi and Duggal (2023) highlight the necessity for scalable innovations and adaptable AI systems to fit existing medical practices. Henry et al. (2022) show that clinicians value ML systems that can adapt to their diagnostic processes, enhancing their decision-making.

Transparent AI Algorithms and Models (Hypothesis 2):
- Prediction: Transparency in AI algorithms and models is crucial for technical strategy, making AI systems easier to understand and use.
- Outcome: Human-centered accountability and AI readiness are enhanced through transparency.
- Case Study Alignment: Henry et al. (2022) demonstrate that clinicians' understanding of ML systems enhances their trust and usability. Morrison (2021) also emphasizes the need for clear communication and standardized language around AI to build trust.

Responsible AI Innovation (Hypothesis 3)
- Prediction: Responsible AI innovation ensures that AI systems are developed with ethical considerations and governance.
- Outcome: Trustworthy AI systems that align with ethical standards.
- Case Study Alignment: Sun (2021) underscores the importance of ethical governance and understanding social power dynamics in AI adoption. Petersson et al. (2022) discuss how responsible AI implementation requires systematic approaches and shared plans.

Human-AI Symbiotic and Synergistic Collaboration (Hypothesis 4)
- Prediction: Collaboration between humans and AI systems enhances the relative advantage of AI, making it a valuable tool for augmenting human capabilities.

- Outcome: Sustainable AI implementation that supports ongoing innovation and collaboration.
- Case Study Alignment: Li et al. (2023) highlight how AI can transform HRM practices through co-creation processes. Ganapathi and Duggal (2023) emphasize the importance of involving doctors in AI development to ensure that AI tools meet clinical needs.

Grounded in practical healthcare case studies, the TOAST framework builds on and extends existing models by incorporating dynamic trust assessments, accountability measures, comprehensive technological, organizational, and environmental considerations, ongoing ethical evaluations, and proactive risk management. These enhancements ensure that AI systems in healthcare are robust, trustworthy, ethically sound, and seamlessly integrated within their operational contexts. The dynamic trust framework posits that trust in AI evolves with technology, user experiences, and societal perceptions. TOAST incorporates mechanisms for regularly evaluating user experiences and AI operations transparency, as Henry et al. highlighted (2022). Regular evaluation ensures AI systems remain trustworthy over time, adapting to technological advancements and changing expectations, thus supporting continuous trust assessment. The AI accountability framework suggests key performance indicators and internal auditing to measure AI's ethical and operational aspects. TOAST extends this by embedding these metrics within corporate governance structures and daily AI operations, as Petersson et al. (2022) show. Regular audits and assessments for ethical compliance enhance trustworthiness and effectiveness, aligning with the accountability framework's goals. The TOE framework assesses technological, organizational, and environmental factors in AI adoption. TOAST integrates these factors by emphasizing technological robustness, organizational readiness, and environmental compliance, as Ganapathi and Duggal (2023) and Dumbach et al. (2021) discussed. This comprehensive approach ensures AI systems are well-suited to their operational contexts and regulatory environments, extending the TOE framework's principles. The ethical AI framework prioritizes continuous ethical investigations to assess AI systems' moral implications. TOAST embeds ethical considerations within its organizational, operational, technical, and reputational pillars, as demonstrated in Sun (2021) and Petersson et al. (2022). Ensuring AI systems align with ethical norms and societal expectations promotes responsible AI practices and builds public trust and confidence. The AI risk framework identifies, analyzes, and mitigates potential risks associated with AI systems. TOAST incorporates this perspective by emphasizing risk mitigation across its pillars, accentuated by Morrison (2021) and Sun (2021). Proactive risk management ensures AI systems contribute constructively to organizational goals and societal welfare, aligning with the AI risk framework's emphasis on comprehensive risk management.

To sum up, the TOAST framework extends the Information Systems (IS) diffusion variance model (Agarwal & Prasad, 1998; Cooper & Zmud, 1990; Crum et al., 1996), which posits that successful IS implementation hinges on technical compatibility, ease of use, and perceived usefulness. TOAST framework adapts these principles to address the specific complexities of AI as an intelligent IS and extends the success factors to denote key strategies that can address the emerging implementation challenges of AI. To support technical compatibility, we need adaptive AI systems development (hypothesis 1). The need for AI systems to adapt to existing infrastructure and practices is evident across the case studies. For example, Dumbach et al. (2021) highlight SMEs' challenges in ensuring AI systems align with regulatory standards and data privacy concerns. To support ease of use, transparent AI algorithms and models (hypothesis 2) are essential technical strategies for responsible AI innovation (hypothesis 3). Morrison (2021) and Henry et al. (2022) show that clear understanding and communication about AI systems are crucial for their acceptance and use. To support perceived usefulness, human-AI symbiotic and synergistic collaboration (hypothesis 4) is crucial for achieving collectively relative advantage. The ability of AI systems to provide a clear benefit over existing practices must be highlighted. For example, Li et al. (2023) and Henry et al. (2022) demonstrate that AI systems need tangible improvements in performance and decision-making processes.

## 6. Discussion

We have introduced a comprehensive AI implementation framework, focused on addressing the unique challenges in healthcare but adaptable across industry sectors, to guide organizations toward the responsible adoption of AI as augmented intelligence.

### 6.1. Contribution to Theories

The framework synergizes theoretical principles with practical insights to ensure a balanced and dynamic approach to sustainable AI implementation. The framework comprises four critical TOAST components: trust and accountability, optimization and control, adaptability and innovation, and socio-technical environment. These dimensions are grounded in established theories—trust theory, control theory, innovation theory, and socio-technical theory, respectively—and address the multifaceted nature of AI implementation, covering technical, ethical, organizational, and innovative aspects essential for successful AI adoption.

An analysis of multi-case studies has demonstrated the practicality and theoretical soundness of the framework within the healthcare sectors, particularly the interconnected interplay between components. Synergizing all TOAST components ensures that organizations can adopt AI tailored to their industry requirements while adhering to best practices and theoretical principles across organizational AI maturity stages. Furthermore, the framework extends existing frameworks and principles vital for the sustainable integration of AI.

### 6.2. Practical Implications

The practical implications of the proposed AI implementation framework are significant, particularly in the case study context of the healthcare sector, where they can profoundly impact patient outcomes, staff efficiency, and organizational adaptability. The implementation framework comprises six phases that aim to establish a foundation for sustainable AI innovation, developing AI applications that enhance clinical workflows, integrate AI systems into existing organizational processes, establish trust and accountability in AI applications, harmonize AI with the cultural fabric of the healthcare sector, and promote ongoing learning and adaptation in AI applications. By following this framework, healthcare organizations can maximize benefits, minimize potential disruptions, and ensure that AI solutions are ethically aligned, culturally sensitive, and strategically implemented.

The selected case studies demonstrates that AI implementation strategies need more than a one-size-fits-all approach. Each case has unique challenges and priorities influenced by its specific operational, cultural, and ethical contexts. Understanding these nuances is key to developing effective strategies for addressing the multifaceted nature of AI adoption across different organizations and contexts. The TOAST framework considers technological readiness, organizational dynamics, cultural sensitivities, and ethical considerations. Building trust among healthcare professionals and addressing ethical and legal concerns are key. Successful AI adoption also involves integrating AI into operational landscapes, considering human-centered AI design, and fostering interdisciplinary collaboration.

### 6.3. Limitations

Successful adoption of AI in various organizations requires tailored strategies considering unique challenges and priorities influenced by operational, cultural, and ethical contexts.

The framework needs continuous evolution to incorporate emerging AI technologies, AI implementation frameworks, and further relevant theories and case studies that address new challenges. Based on long-term analysis of longitudinal studies, the AI implementation framework can be refined to be a more robust, adaptable, and comprehensive guide for organizations aiming to navigate the complexities of AI adoption in an ethical, socially responsible, and innovative manner. Future work should also continue to refine the interplay between the framework's dimensions and implementation phases to ensure that AI adoption is comprehensive and addresses technical, ethical, and cultural aspects. By aligning the framework's dimensions and phases with the specific needs and challenges of different industry sectors, organizations can leverage AI as a powerful tool for augmented intelligence. Organizations must ensure AI solutions are effective and efficient but also responsible, ethical, and adaptable to organizational changes.

## 7. Conclusion

This paper provides a comprehensive theoretical framework that elucidates the complexities and challenges of adopting AI in organizations. The framework offers a holistic approach to managing the interplay between humans and AI systems by integrating trust, control, socio-technical, and innovation theories. This approach ensures that AI adoption aligns with ethical standards, promotes social congruity, advances innovation, and promotes a sustainable implementation of AI in business.

Our healthcare case studies analysis underscores the framework's utility in real-world scenarios. It highlights the framework's adaptability across different organizational contexts and its potential to guide the systematic implementation of AI technologies. The framework fosters a deeper understanding of the socio-technical dynamics at play, assisting organizations in navigating the intricacies of AI integration. It enhances trust, optimizes collaborative efforts, and maintains a continuous innovation cycle.

Future research should focus on refining the framework by incorporating emerging technologies and ongoing feedback from practical applications. These efforts will keep the framework relevant and effective in the face of rapidly evolving AI capabilities and increasingly complex organizational environments. For practitioners, this study offers a structured pathway for the phased implementation of AI. It fosters robust, responsible, and innovative organizational practices that can adapt to future technological advancements and market demands.

# References

Abonamah, A. A., & Abdelhamid, N. (2024). Managerial insights for AI/ML implementation: A playbook for successful organizational integration. *Discover Artificial Intelligence*, *4*(1), 22. https://doi.org/10.1007/s44163-023-00100-5

Adadi, A., & Berrada, M. (2018). Peeking Inside the Black-Box: A Survey on Explainable Artificial Intelligence (XAI). *IEEE Access*, *6*, 52138–52160. https://doi.org/10.1109/ACCESS.2018.2870052

Adewuyi, A. A., Cheng, H., Shi, Q., Cao, J., MacDermott, Á., & Wang, X. (2019). CTRUST: A Dynamic Trust Model for Collaborative Applications in the Internet of Things. *IEEE Internet of Things Journal*, *6*(3), 5432–5445. https://doi.org/10.1109/JIOT.2019.2902022

Agarwal, R., & Prasad, J. (1998). A Conceptual and Operational Definition of Personal Innovativeness in the Domain of Information Technology. *Information Systems Research*, *9*(2), 204–215.

Alhashmi, S. F. S., Alshurideh, M., Al Kurdi, B., & Salloum, S. A. (2020). A Systematic Review of the Factors Affecting the Artificial Intelligence Implementation in the Health Care Sector. In A.-E. Hassanien, A. T. Azar, T. Gaber, D. Oliva, & F. M. Tolba (Eds.), *Proceedings of the International Conference on Artificial Intelligence and Computer Vision (AICV2020)* (pp. 37–49). Springer International Publishing. https://doi.org/10.1007/978-3-030-44289-7_4

Alsheibani, S., Cheung, Y., & Messom, C. (2018). *Artificial Intelligence Adoption: AI-readiness at Firm-Level*.

Anderson, C., Baskerville, R. L., & Kaul, M. (2017). Information Security Control Theory: Achieving a Sustainable Reconciliation Between Sharing and Protecting the Privacy of Information. *Journal of Management Information Systems*, *34*(4), 1082–1112. https://doi.org/10.1080/07421222.2017.1394063

Ångström, R. C., Björn, M., Dahlander, L., Mähring, M., & Wallin, M. W. (2023). Getting AI Implementation Right: Insights from a Global Survey. *California Management Review*, *66*(1), 5–22. https://doi.org/10.1177/00081256231190430

Araujo, T., Helberger, N., Kruikemeier, S., & de Vreese, C. H. (2020). In AI we trust? Perceptions about automated decision-making by artificial intelligence. *AI & SOCIETY*, *35*(3), 611–623. https://doi.org/10.1007/s00146-019-00931-w

Arias-Pérez, J., & Vélez-Jaramillo, J. (2022). Ignoring the three-way interaction of digital orientation, Not-invented-here syndrome and employee's artificial intelligence awareness in digital innovation performance: A recipe for failure. *Technological Forecasting and Social Change*, *174*, 121305. https://doi.org/10.1016/j.techfore.2021.121305

Asatiani, A., Malo, P., Nagbøl, P., Penttinen, E., Rinta-Kahila, T., & Salovaara, A. (2021). Sociotechnical Envelopment of Artificial Intelligence: An Approach to Organizational Deployment of Inscrutable Artificial Intelligence Systems. *Journal of the Association for Information Systems*, *22*(2). https://doi.org/10.17705/1jais.00664

Attard-Frost, B., De los Ríos, A., & Walters, D. R. (2023). The ethics of AI business practices: A review of 47 AI ethics guidelines. *AI and Ethics*, *3*(2), 389–406. https://doi.org/10.1007/s43681-022-00156-6

Bankins, S., Ocampo, A. C., Marrone, M., Restubog, S. L. D., & Woo, S. E. (2023). A multilevel review of artificial intelligence in organizations: Implications for organizational behavior research and practice. *Journal of Organizational Behavior*, *n/a*(n/a). https://doi.org/10.1002/job.2735

Baxter, G., & Sommerville, I. (2011). Socio-technical systems: From design methods to systems engineering. *Interacting with Computers*, *23*(1), 4–17. https://doi.org/10.1016/j.intcom.2010.07.003

Borges, A. F. S., Laurindo, F. J. B., Spínola, M. M., Gonçalves, R. F., & Mattos, C. A. (2021). The strategic use of artificial intelligence in the digital era: Systematic literature review and future research directions. *International Journal of Information Management*, *57*, 102225. https://doi.org/10.1016/j.ijinfomgt.2020.102225

Brock, J. K.-U., & Von Wangenheim, F. (2019). Demystifying AI: What digital transformation leaders can teach you about realistic artificial intelligence. *California Management Review*, *61*(4), 110–134.

Burström, T., Parida, V., Lahti, T., & Wincent, J. (2021). AI-enabled business-model innovation and transformation in industrial ecosystems: A framework, model and outline for further research. *Journal of Business Research*, *127*, 85–95.

Cabiddu, F., Moi, L., Patriotta, G., & Allen, D. G. (2022). Why do users trust algorithms? A review and conceptualization of initial trust and trust over time. *European Management Journal*, *40*(5), 685–706. https://doi.org/10.1016/j.emj.2022.06.001

Caldas, R. D., Rodrigues, A., Gil, E. B., Rodrigues, G. N., Vogel, T., & Pelliccione, P. (2020). A hybrid approach combining control theory and AI for engineering self-adaptive systems. *Proceedings of the IEEE/ACM 15th International Symposium on Software Engineering for Adaptive and Self-Managing Systems*, 9–19. https://doi.org/10.1145/3387939.3391595

Chatila, R., Dignum, V., Fisher, M., Giannotti, F., Morik, K., Russell, S., & Yeung, K. (2021). Trustworthy AI. In B. Braunschweig & M. Ghallab (Eds.), *Reflections on Artificial Intelligence for Humanity* (pp. 13–39). Springer International Publishing. https://doi.org/10.1007/978-3-030-69128-8_2

Chatterjee, D., & Rao, S. (2020). Computational Sustainability: A Socio-technical Perspective. *ACM Computing Surveys*, *53*(5), 101:1-101:29. https://doi.org/10.1145/3409797

Cho, J.-H., Chan, K., & Adali, S. (2015). A Survey on Trust Modeling. *ACM Computing Surveys*, *48*(2), 28:1-28:40. https://doi.org/10.1145/2815595

Chowdhury, S., Budhwar, P., Dey, P. K., Joel-Edgar, S., & Abadie, A. (2022). AI-employee collaboration and business performance: Integrating knowledge-based view, socio-technical systems and organisational socialisation framework. *Journal of Business Research*, *144*, 31–49. https://doi.org/10.1016/j.jbusres.2022.01.069

Clemmensen, T. (2021). *Human Work Interaction Design: A Platform for Theory and Action*. Springer International Publishing. https://doi.org/10.1007/978-3-030-71796-4

Cooper, R. B., & Zmud, R. W. (1990). Information Technology Implementation Research: A Technological Diffusion Approach. *Management Science*, *36*(2), 123–139. https://doi.org/10.1287/mnsc.36.2.123

Crum, M. R., Premkumar, G., & Ramamurthy, K. (1996). An Assessment of Motor Carrier Adoption, Use, and Satisfaction with EDI. *Transportation Journal*, *35*(4), 44–57.

Di Vaio, A., Palladino, R., Hassan, R., & Escobar, O. (2020). Artificial intelligence and business models in the sustainable development goals perspective: A systematic literature review. *Journal of Business Research*, *121*, 283–314. https://doi.org/10.1016/j.jbusres.2020.08.019

Dignum, V. (2019). *Responsible Artificial Intelligence: How to Develop and Use AI in a Responsible Way*. Springer Nature. https://doi.org/10.1007/978-3-030-30371-6.

Duan, Y., Edwards, J. S., & Dwivedi, Y. K. (2019). Artificial intelligence for decision making in the era of Big Data – evolution, challenges and research agenda. *International Journal of Information Management*, *48*, 63–71. https://doi.org/10.1016/j.ijinfomgt.2019.01.021

Dumbach, P., Liu, R., Jalowski, M., & Eskofier, B. (2021). *The Adoption Of Artificial Intelligence In SMEs—A Cross-National Comparison In German And Chinese Healthcare*. BIR Workshops. https://www.semanticscholar.org/paper/The-Adoption-Of-Artificial-Intelligence-In-SMEs-A-Dumbach-Liu/9b48618859a4751c682dd8e4e72afd329ca00d54

Dwivedi, Y. K., Hughes, L., Ismagilova, E., Aarts, G., Coombs, C., Crick, T., Duan, Y., Dwivedi, R., Edwards, J., Eirug, A., Galanos, V., Ilavarasan, P. V., Janssen, M., Jones, P., Kar, A. K., Kizgin, H., Kronemann, B., Lal, B., Lucini, B., … Williams, M. D. (2021). Artificial Intelligence (AI): Multidisciplinary perspectives on emerging challenges, opportunities, and agenda for research, practice and policy. *International Journal of Information Management*, *57*, 101994. https://doi.org/10.1016/j.ijinfomgt.2019.08.002

Eitel-Porter, R., Corcoran, M., & Connolly, P. (2021). *Responsible AI Principles to Practice* [Research Report]. Accenture. https://www.accenture.com/au/insights/artificial-intelligence/responsible-ai-principles-practice

Enholm, I. M., Papagiannidis, E., Mikalef, P., & Krogstie, J. (2022). Artificial Intelligence and Business Value: A Literature Review. *Information Systems Frontiers*, *24*(5), 1709–1734. https://doi.org/10.1007/s10796-021-10186-w

Ferrario, A., Loi, M., & Viganò, E. (2020). In AI We Trust Incrementally: A Multi-layer Model of Trust to Analyze Human-Artificial Intelligence Interactions. *Philosophy & Technology*, *33*(3), 523–539. https://doi.org/10.1007/s13347-019-00378-3

Filieri, A., Maggio, M., Angelopoulos, K., D'Ippolito, N., Gerostathopoulos, I., Hempel, A. B., Hoffmann, H., Jamshidi, P., Kalyvianaki, E., Klein, C., Krikava, F., Misailovic, S., Papadopoulos, A. V., Ray, S., Sharifloo, A. M., Shevtsov, S., Ujma, M., & Vogel, T. (2015). Software Engineering Meets Control Theory. *2015 IEEE/ACM 10th International Symposium on Software Engineering for Adaptive and Self-Managing Systems*, 71–82. https://doi.org/10.1109/SEAMS.2015.12

Floridi, L., Cowls, J., Beltrametti, M., Chatila, R., Chazerand, P., Dignum, V., Luetge, C., Madelin, R., Pagallo, U., Rossi, F., Schafer, B., Valcke, P., & Vayena, E. (2021). An Ethical Framework for a Good AI Society: Opportunities, Risks, Principles, and Recommendations. In L. Floridi (Ed.), *Ethics, Governance, and Policies in Artificial Intelligence* (pp. 19–39). Springer International Publishing. https://doi.org/10.1007/978-3-030-81907-1_3

Furman, J., & Seamans, R. (2019). AI and the Economy. *Innovation Policy and the Economy*, *19*(1), 161–191.

Ganapathi, S., & Duggal, S. (2023). Exploring the experiences and views of doctors working with Artificial Intelligence in English healthcare; a qualitative study. *PLOS ONE*, *18*(3), e0282415. https://doi.org/10.1371/journal.pone.0282415

Gill, S. S., Xu, M., Ottaviani, C., Patros, P., Bahsoon, R., Shaghaghi, A., Golec, M., Stankovski, V., Wu, H., Abraham, A., Singh, M., Mehta, H., Ghosh, S. K., Baker, T., Parlikad, A. K., Lutfiyya, H., Kanhere, S. S., Sakellariou, R., Dustdar, S., … Uhlig, S. (2022). AI for next generation computing: Emerging trends and future directions. *Internet of Things*, *19*, 100514. https://doi.org/10.1016/j.iot.2022.100514

Gille, F., Jobin, A., & Ienca, M. (2020). What we talk about when we talk about trust: Theory of trust for AI in healthcare. *Intelligence-Based Medicine*, *1–2*, 100001. https://doi.org/10.1016/j.ibmed.2020.100001

Glikson, E., & Woolley, A. W. (2020). Human Trust in Artificial Intelligence: Review of Empirical Research. *Academy of Management Annals*, *14*(2), 627–660. https://doi.org/10.5465/annals.2018.0057

Haefner, N., Wincent, J., Parida, V., & Gassmann, O. (2021). Artificial intelligence and innovation management: A review, framework, and research agenda☆. *Technological Forecasting and Social Change*, *162*, 120392. https://doi.org/10.1016/j.techfore.2020.120392

Henry, K. E., Kornfield, R., Sridharan, A., Linton, R. C., Groh, C., Wang, T., Wu, A., Mutlu, B., & Saria, S. (2022). Human–machine teaming is key to AI adoption: Clinicians' experiences with a deployed machine learning system. *Npj Digital Medicine*, *5*(1), Article 1. https://doi.org/10.1038/s41746-022-00597-7

Herrmann, T., & Pfeiffer, S. (2023). Keeping the organization in the loop: A socio-technical extension of human-centered artificial intelligence. *AI & SOCIETY*, *38*(4), 1523–1542. https://doi.org/10.1007/s00146-022-01391-5

Hevner, A., & Storey, V. (2023). Research Challenges for the Design of Human-Artificial Intelligence Systems (HAIS). *ACM Transactions on Management Information Systems*, *14*(1), 10:1-10:18. https://doi.org/10.1145/3549547

Hoda, R. (2022). Socio-Technical Grounded Theory for Software Engineering. *IEEE Transactions on Software Engineering*, *48*(10), 3808–3832. https://doi.org/10.1109/TSE.2021.3106280

Holton, R., & Boyd, R. (2021). 'Where are the people? What are they doing? Why are they doing it?'(Mindell) Situating artificial intelligence within a socio-technical framework. *Journal of Sociology*, *57*(2), 179–195. https://doi.org/10.1177/1440783319873046

Hou, Z.-S., & Wang, Z. (2013). From model-based control to data-driven control: Survey, classification and perspective. *Information Sciences*, *235*, 3–35. https://doi.org/10.1016/j.ins.2012.07.014

Jacovi, A., Marasović, A., Miller, T., & Goldberg, Y. (2021). Formalizing Trust in Artificial Intelligence: Prerequisites, Causes and Goals of Human Trust in AI. *Proceedings of the 2021 ACM Conference on Fairness, Accountability, and Transparency*, 624–635. https://doi.org/10.1145/3442188.3445923

Jarrahi, M. H. (2018). Artificial intelligence and the future of work: Human-AI symbiosis in organizational decision making. *Business Horizons*, *61*(4), 577–586. https://doi.org/10.1016/j.bushor.2018.03.007

Jöhnk, J., Weißert, M., & Wyrtki, K. (2021). Ready or not, AI comes—An interview study of organizational AI readiness factors. *Business & Information Systems Engineering*, *63*, 5–20.

Khargonekar, P. P., & Dahleh, M. A. (2018). Advancing systems and control research in the era of ML and AI. *Annual Reviews in Control*, *45*, 1–4. https://doi.org/10.1016/j.arcontrol.2018.04.001

Li, P., Bastone, A., Mohamad, T. A., & Schiavone, F. (2023). How does artificial intelligence impact human resources performance. Evidence from a healthcare institution in the United Arab Emirates. *Journal of Innovation & Knowledge*, *8*(2). https://doi.org/10.1016/j.jik.2023.100340

Liu, B. (2021). In AI We Trust? Effects of Agency Locus and Transparency on Uncertainty Reduction in Human–AI Interaction. *Journal of Computer-Mediated Communication*, *26*(6), 384–402. https://doi.org/10.1093/jcmc/zmab013

Loureiro, S. M. C., Guerreiro, J., & Tussyadiah, I. (2021). Artificial intelligence in business: State of the art and future research agenda. *Journal of Business Research*, *129*, 911–926. https://doi.org/10.1016/j.jbusres.2020.11.001

Mariani, M. M., Machado, I., Magrelli, V., & Dwivedi, Y. K. (2023). Artificial intelligence in innovation research: A systematic review, conceptual framework, and future research directions. *Technovation*, *122*, 102623. https://doi.org/10.1016/j.technovation.2022.102623

Maslej, N., Fattorini, L., Brynjolfsson, E., Etchemendy, J., Ligett, K., Lyons, T., Manyika, J., Ngo, H., Niebles, J. C., Parli, V., Shoham, Y., Wald, R., Clark, J., & Perrault, R. (2023). *The AI Index 2023 Annual Report*. AI Index Steering Committee, Institute for Human-Centered AI, Stanford University.

Morrison, K. (2021). Artificial intelligence and the NHS: A qualitative exploration of the factors influencing adoption. *Future Healthc J*, *8*(3), e648–e654. https://doi.org/10.7861/fhj.2020-0258

Mueller, S. T., Hoffman, R. R., Clancey, W., Emrey, A., & Klein, G. (2019). Explanation in human-AI systems: A literature meta-review, synopsis of key ideas and publications, and bibliography for explainable AI. *arXiv Preprint arXiv:1902.01876*.

Neumann, O., Guirguis, K., & Steiner, R. (2022). Exploring artificial intelligence adoption in public organizations: A comparative case study. *Public Management Review*, 1–28. https://doi.org/10.1080/14719037.2022.2048685

Omrani, N., Rivieccio, G., Fiore, U., Schiavone, F., & Agreda, S. G. (2022). To trust or not to trust? An assessment of trust in AI-based systems: Concerns, ethics and contexts. *Technological Forecasting and Social Change*, *181*, 121763. https://doi.org/10.1016/j.techfore.2022.121763

Petersson, L., Larsson, I., Nygren, J. M., Nilsen, P., Neher, M., Reed, J. E., Tyskbo, D., & Svedberg, P. (2022). Challenges to implementing artificial intelligence in healthcare: A qualitative interview study with healthcare leaders in Sweden. *BMC Health Services Research*, *22*(1), 850. https://doi.org/10.1186/s12913-022-08215-8

Prem, E. (2023). From ethical AI frameworks to tools: A review of approaches. *AI and Ethics*, *3*(3), 699–716. https://doi.org/10.1007/s43681-023-00258-9

Raji, I. D., Smart, A., White, R. N., Mitchell, M., Gebru, T., Hutchinson, B., Smith-Loud, J., Theron, D., & Barnes, P. (2020). Closing the AI accountability gap: Defining an end-to-end framework for internal algorithmic auditing. *Proceedings of the 2020 Conference on Fairness, Accountability, and Transparency*, 33–44. https://doi.org/10.1145/3351095.3372873

Ransbotham, S. (2017). *Reshaping business with artificial intelligence: Closing the gap between ambition and action"*. https://sloanreview.mit.edu/projects/reshaping-business-with-artificial-intelligence/

Razmerita, L., Brun, A., & Nabeth, T. (2022). Collaboration in the Machine Age: Trustworthy Human-AI Collaboration. In M. Virvou, G. A. Tsihrintzis, & L. C. Jain (Eds.), *Advances in Selected Artificial Intelligence Areas: World Outstanding Women in Artificial Intelligence* (pp. 333–356). Springer International Publishing. https://doi.org/10.1007/978-3-030-93052-3_14

Robinson, S. C. (2020). Trust, transparency, and openness: How inclusion of cultural values shapes Nordic national public policy strategies for artificial intelligence (AI). *Technology in Society*, *63*, 101421. https://doi.org/10.1016/j.techsoc.2020.101421

Sartori, L., & Bocca, G. (2023). Minding the gap(s): Public perceptions of AI and socio-technical imaginaries. *AI & SOCIETY*, *38*(2), 443–458. https://doi.org/10.1007/s00146-022-01422-1

Sartori, L., & Theodorou, A. (2022). A sociotechnical perspective for the future of AI: Narratives, inequalities, and human control. *Ethics and Information Technology*, *24*(1), 4. https://doi.org/10.1007/s10676-022-09624-3

Seok, H., Nof, S. Y., & Filip, F. G. (2012). Sustainability decision support system based on collaborative control theory. *Annual Reviews in Control*, *36*(1), 85–100. https://doi.org/10.1016/j.arcontrol.2012.03.007

Shneiderman, B. (2020). Human-centered artificial intelligence: Reliable, safe & trustworthy. *International Journal of Human–Computer Interaction*, *36*(6), 495–504.

Soma, R., Bratteteig, T., Saplacan, D., Schimmer, R., Campano, E., & Verne, G. (2022). Strengthening Human Autonomy. In the era of autonomous technology. *Scandinavian Journal of Information Systems*, *34*(2). https://aisel.aisnet.org/sjis/vol34/iss2/5

Sony, M., & Naik, S. (2020). Industry 4.0 integration with socio-technical systems theory: A systematic review and proposed theoretical model. *Technology in Society*, *61*, 101248. https://doi.org/10.1016/j.techsoc.2020.101248

Tabassi, E. (2023). Artificial Intelligence Risk Management Framework (AI RMF 1.0). *NIST*. https://www.nist.gov/publications/artificial-intelligence-risk-management-framework-ai-rmf-10

Toreini, E., Aitken, M., Coopamootoo, K. P. L., Elliott, K., Zelaya, V. G., Missier, P., Ng, M., & van Moorsel, A. (2022). *Technologies for Trustworthy Machine Learning: A Survey in a Socio-Technical Context* (arXiv:2007.08911). arXiv. https://doi.org/10.48550/arXiv.2007.08911

Tóth, Z., Caruana, R., Gruber, T., & Loebbecke, C. (2022). The dawn of the AI Robots: Towards a new framework of AI robot accountability. *Journal of Business Ethics*, *178*(4), 895–916.

Uren, V., & Edwards, J. S. (2023). Technology readiness and the organizational journey towards AI adoption: An empirical study. *International Journal of Information Management*, *68*, 102588. https://doi.org/10.1016/j.ijinfomgt.2022.102588

Vassilakopoulou, P. (2020). *Sociotechnical approach for accountability by design in AI systems*.

Vereschak, O., Bailly, G., & Caramiaux, B. (2021). How to Evaluate Trust in AI-Assisted Decision Making? A Survey of Empirical Methodologies. *Proceedings of the ACM on Human-Computer Interaction*, *5*(CSCW2), 1–39. https://doi.org/10.1145/3476068

Verganti, R., Vendraminelli, L., & Iansiti, M. (2020). Innovation and Design in the Age of Artificial Intelligence. *Journal of Product Innovation Management*, *37*(3), 212–227. https://doi.org/10.1111/jpim.12523

Wirtz, B. W., Weyerer, J. C., & Kehl, I. (2022). Governance of artificial intelligence: A risk and guideline-based integrative framework. *Government Information Quarterly*, *39*(4), 101685. https://doi.org/10.1016/j.giq.2022.101685

Yu, X., Xu, S., & Ashton, M. (2022). Antecedents and outcomes of artificial intelligence adoption and application in the workplace: The socio-technical system theory perspective. *Information Technology & People*, *36*(1), 454–474. https://doi.org/10.1108/ITP-04-2021-0254

Zimmer, M. P., Minkkinen, M., & Mäntymäki, M. (2022). *Responsible Artificial Intelligence Systems Critical considerations for business model design*. *34*.